# A NOVEL SCHEME TO IMPROVE THE SPECTRUM SENSING PERFORMANCE

S.Taruna[1] and Bhumika Pahwa[2]

[1]Computer Science Department, Banasthali Vidyapith, Jaipur, Rajasthan
[2]Information Technology Department, Banasthali Vidyapith, Jaipur, Rajasthan

***ABSTRACT***

*Due to limited availability of spectrum for licensed users only, the need for secondary access by unlicensed users is increasing. Cognitive radio turns out to be helping this situation because all that is needed is a technique that could efficiently detect the empty spaces and provide them to the secondary devices without causing any interference to the primary (licensed) users. Spectrum sensing is the foremost function of the cognitive radio which senses the environment for white spaces. Energy detection is one of the various spectrum sensing techniques that are under research. Earlier it was shown that energy detection works better under AWGN channel as compared to Rayleigh channel, however the conventional spectrum sensing techniques have a high probability of false alarm and also show a better probability of detection for higher values of SNR. There is a need for a new technique that shows a reduced probability of false alarm as well as an increase in the probability of detection for lower values of SNR. In the present work the conventional energy detection technique has been enhanced to get better results.*

## 1. INTRODUCTION

With the ever rising wireless communication, setbacks like bandwidth scarcity have gained more attention. While, the recent studies by FCC have shown that huge portion of spectrum is vacant most of the time. This portion is the licensed spectrum band which can only be utilized by licensed users only. Hence, to resolve this problem of under-utilized spectrum, secondary users are permitted by the FCC to utilize the licensed band when it is not in use and named it as cognitive radio. Spectrum sensing is used to sense the presence of licensed users [1].Energy Detection, Matched Filter detection and Cyclostationary detection are the three standard methods used for spectrum sensing. All these three techniques have been explained further. Section 2 explains the various wireless channels; section 3 throws light on the spectrum sensing definition and various probabilities that help us realize the results. Section 4 explains the energy detection spectrum sensing the Section 5 shows the proposed method and simulation results while Section 6 concludes the paper.

## 2. WIRELESS CHANNELS

Wireless channels provide the physical means to transport a signal produced by the transmitter and delivers it to the receiver. Behaviour of a wireless communication channel varies with time. On the other hand, models used to describe the behaviour of wireless links are not precise; they may offer a rough calculation. To portray wireless channel behaviour various models are available, such as an AWGN (Additive White Gaussian Noise) model and fading model like Rayleigh model.

### 2.1. AWGN Channel

A straight forward environment for a wireless communication system to operate is the AWGN environment. Due to the channel effect, the transmitted channel is summated with a random signal. Henceforth, the received signal s(t), can be expressed as-

$$s(t) = x(t) + n(t) \qquad (1)$$

Where, x(t) is the transmitted signal, and n(t) is the background signal.[2]

### 2.2. Rayleigh Fading Channel

A communication channel that faces different fading phenomenon during signal transmission is called a fading channel. Multipath propagation is the major cause of fading. The signal that arrives at the receiver comes from different paths having different delays and path gains.

These propagations paths might seem to be destructive or constructive. Finally the received signal is the algebraic sum of various paths of propagation so that some of the paths are added and the other are subtracted.
Primary reason for Rayleigh fading is the multipath reception of the transmitted signal. There is no direct line of sight between the receiver and the transmitter.
Consider a transmitted signal, x(t),

$$x(t) = \cos(\omega_c t) \qquad (2)$$

Where, $\omega_c$ is the transmitted signal frequency in radian/sec, hence, received signal, s(t), can be expressed as-

$$s(t) = \sum a_i \cos(\omega_c t + \phi_i) \qquad (3)$$

where, N is the number of paths, $\phi_i$ is the phase shift of each path, that depends on delay difference and takes values 0 to 2π, and $a_i$ is the amplitude of each path *i*[2].

## 3. Spectrum Sensing

The task of sensing the radio spectrum in its local neighbourhood of the cognitive radio, to find the spectrum holes is defined as spectrum sensing [3]. Due to various reasons spectrum sensing is considered as the most essential function of the cognitive radio.

Basic duty of spectrum sensing is to identify the spectrum white spaces i.e. spectrums which are currently unused by the primary users. Then it can broadcast in these spaces in a way that no interference is caused to the primary users. Another motive of cognitive radio to sense the wireless medium is to sense if any other secondary device is transmitting. Here, the cognitive radio needs to share either some or all the channels occupied by other secondary devices with the object of reducing its own blocking probability. [4][7][8]

Spectrum sensing problem can be put together as a binary hypothesis testing, with two hypotheses –

$$H_o: y[n] = w[n]; n= 1, 2... N$$
$$H_1: y[n] = x[n] + w[n]; n= 1, 2... N$$

Where, $H_o$ states that received signal samples y[n] correspond to noise sample signal w[n] and hence, primary signal is not sensed to be present in the spectrum band. $H_1$ indicates the presence of some primary users signal x[n]. N denotes the number of samples gathered during sensing period.

Ideally the spectrum sensor would select $H_1$ to show the presence of primary users and $H_o$ otherwise. In practice spectrum sensing algorithms fall into mistakes, which are classified as missed detection and false-alarm, which may be defined as-

**Probability of Missed detection, $P_{MD}$**

The condition when a primary user is detected to be absent while it is actually present, is called the probability of missed detection. Higher value of $P_{MD}$ leads to higher interference because in this case the secondary user will assume that the spectrum is free while the spectrum is actually utilized by the primary users.

$$P_{MD} = P(H_O / H_1) \quad (4)$$

**Probability of detection, $P_D$**

The probability of detection is the condition when the primary users are detected to be present while they are actually present, higher value of $P_D$ avoids any interference from the secondary users if they are trying to access the spectrum. A high value of $P_D$ will lead to efficient use of the spectrum without causing interference to the primary user [5].

$$P_D = P(H_1 / H_1) \quad (5), \text{ or}$$

$$P_D = 1 - P_{MD} \quad (6)$$

**Probability of False alarm, $P_{FA}$**

It is defined as the probability of detecting that primary user is present while it is actually absent, and this leads to inefficient utilization of the spectrum. Because, even if the spectrum is free, the secondary user will assume that it is occupied by the primary user and hence will not be able to utilize the spectrum. A low value of $P_{FA}$ is expected to increase the channel reuse capability when it is free [5].

$$P_{FA} = P(H_1 / H_0) \quad (7)$$

Current literature for spectrum sensing is yet in its premature stages of development. Various methods are proposed to discover the presence of signal transmissions. In these approaches, the features of the identified transmission are detected for deciding the signal transmission as well as the signal type. Some of the commonly used spectrum sensing techniques in cognitive radio is explained in this section.

## 4. ENERGY DETECTION SPECTRUM SENSING

Energy detector based technique also known as periodogram, is the most commonly used technique for the purpose of spectrum sensing due to its low implementation and computational complexities [9]. The benefit is that it does not need any information about the primary user signal, hence when the receiver is unable to gather enough information about the primary signal then this technique is the most appropriate to use.[4][6]

To determine the energy of the received signal, output signal from band-pass filter with bandwidth W is squared and integrated over the observation interval T.

Now finally, Y, the output of the integrator is compared with a threshold λ to decide if a licensed user is present or not [10]. Only shortcoming of this method is the increased sensing time. Major challenges of energy detector based sensing are the selection of threshold to detect the primary users, the inability to differentiate whether the interference being caused is from primary users or noise, and its poor performance when the Signal-to-Noise ratio is low.

Hence, energy detector is considered to be optimal when the cognitive devices have no prior information about the features of primary signals except local noise statistics [11].

Energy Detector is composed of four main components:

1) Pre-filter.
2) A/D Converter (Analog to Digital Converter).
3) Squaring Device.
4) Integrator.

Output of the integrator is the energy of the filtered received signal over the time interval $T$ and this output is considered the test statistic to test the two Hypotheses $H_0$ and $H_1$[12].

$H_0$ : corresponds to the absence of the signal and only presence of noise.
$H_1$ : corresponds to the presence of both signal and noise.
Figure 1, shows the block diagram of energy detector.

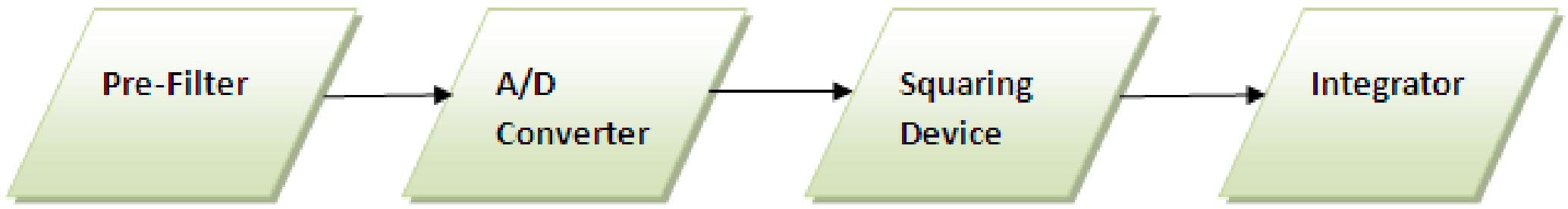

Figure 1: Block diagram of Energy Detector

The results below show that energy detection performs better in AWGN channel when compared with Rayleigh channel.[13]

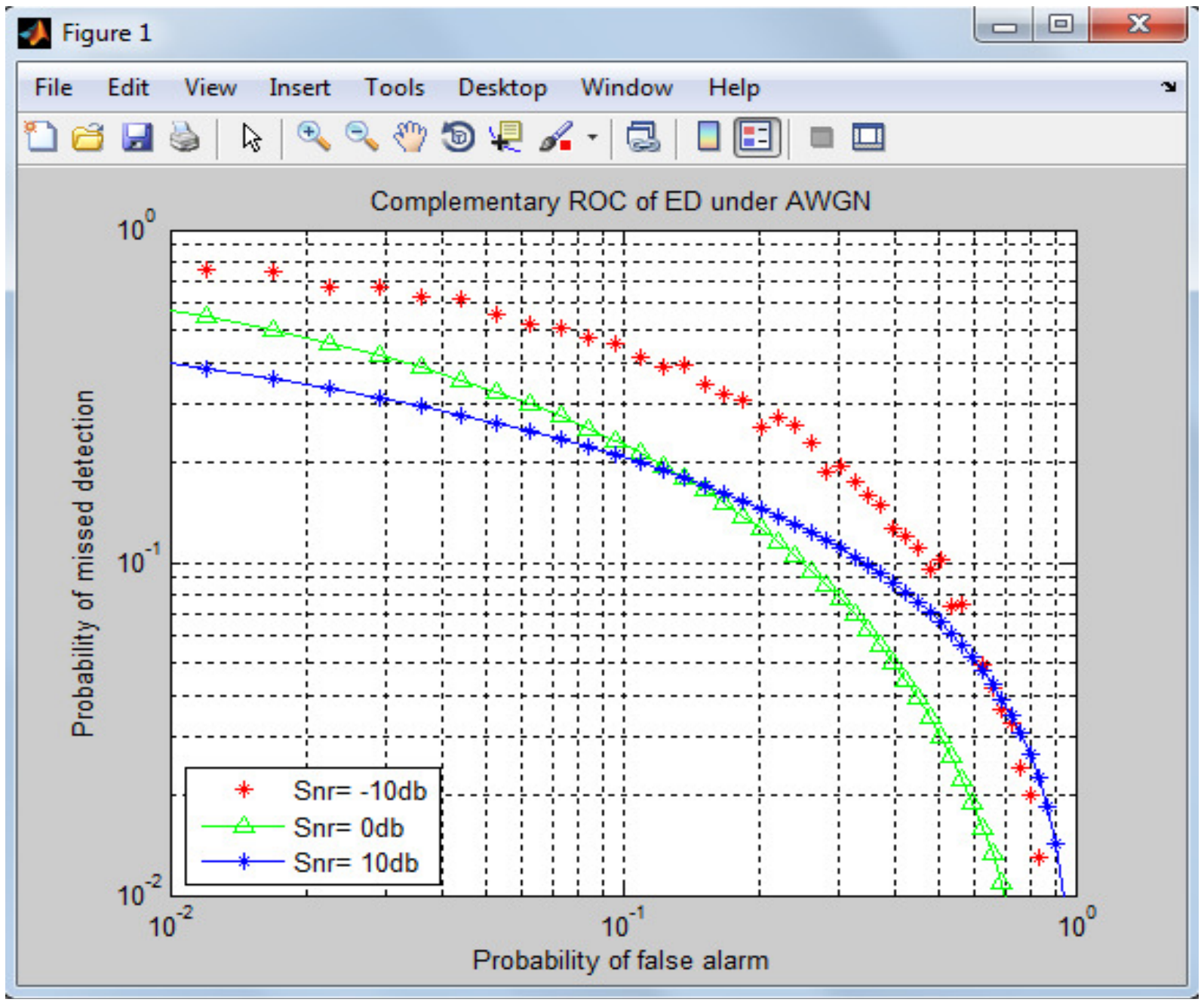

Figure 2: ROC for Energy detection technique under AWGN channels.

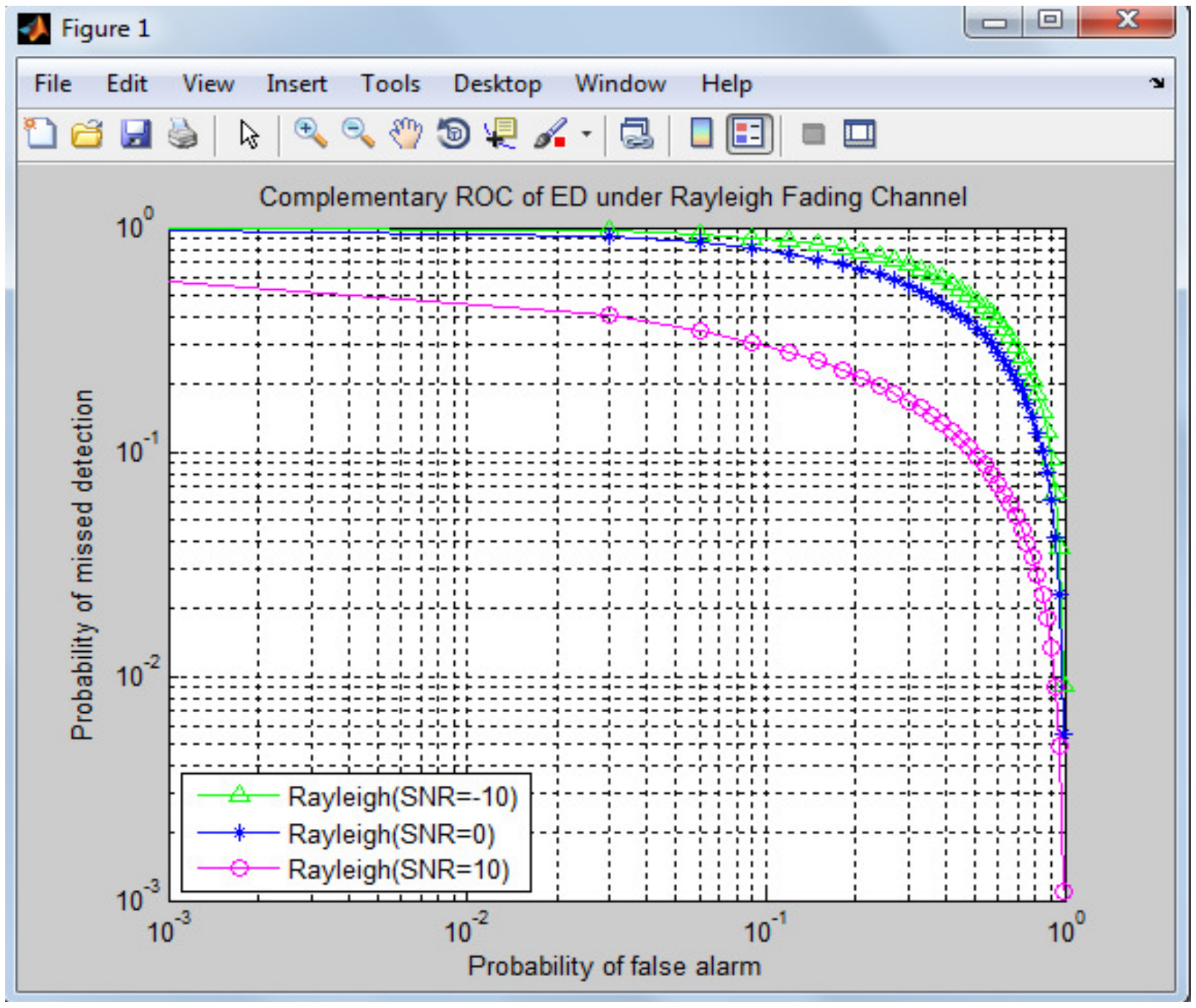

Figure 3: ROC for Energy detection under Rayleigh channels.

Graph in figure 6 shows that probability of missed detection decreases as we replaced squaring operation with cubing operation, a prominent change is seen in probability of missed detection especially for the negative value of SNR, it is also noted from the graph that the probability of false alarm evidently decreases, which shows that using the cubing method instead of squaring leads to a much raised performance of spectrum sensing.

When both the methods are compared it is derived that the conventional energy detection method has a low performance as compared to the improved technique.

The following tables show the results on which the above graphs are based.

Table1: $P_{MD}$ for conventional energy detector

| SNR → / $P_{MD}$ ↓ | -10 db | 0 db | 10 db |
|---|---|---|---|
| 1 | 0.9690 | 0.9260 | 0.7851 |
| 2 | 0.9170 | 0.8162 | 0.6851 |
| 3 | 0.8800 | 0.7309 | 0.5649 |
| 4 | 0.8110 | 0.6601 | 0.4983 |
| 5 | 0.7300 | 0.5994 | 0.3827 |
| 6 | 0.6640 | 0.5463 | 0.3328 |
| 7 | 0.6280 | 0.4573 | 0.2938 |
| 8 | 0.5950 | 0.4194 | 0.2616 |
| 9 | 0.4930 | 0.3851 | 0.1994 |
| 10 | 0.3950 | 0.3539 | 0.1795 |
| 11 | 0.3290 | 0.2991 | 0.1615 |
| 12 | 0.2580 | 0.2324 | 0.1532 |
| 13 | 0.2140 | 0.1960 | 0.1453 |
| 14 | 0.2070 | 0.1647 | 0.1304 |
| 15 | 0.1790 | 0.0947 | 0.1041 |
| 16 | 0.1440 | 0.0774 | 0.0867 |
| 17 | 0.1210 | 0.0561 | 0.0760 |
| 18 | 0.1080 | 0.0392 | 0.0565 |
| 19 | 0.0890 | 0.0260 | 0.0432 |
| 20 | 0.0780 | 0.0190 | 0.0329 |
| 21 | 0.0600 | 0.0133 | 0.0265 |
| 22 | 0.0450 | 0.0109 | 0.0184 |
| 23 | 0.0380 | 0.0068 | 0.0142 |
| 24 | 0.0230 | 0.0037 | 0.0098 |
| 25 | 0.0170 | 0.0024 | 0.0047 |
| 26 | 0.0080 | 0.0015 | 0.0020 |

Table 2: $P_{MD}$ for improved energy detector

| SNR → / $P_{MD}$ ↓ | -10 db | 0 db | 10 db |
|---|---|---|---|
| 1 | 0.6750 | 0.6473 | 0.7776 |
| 2 | 0.6340 | 0.6229 | 0.7402 |
| 3 | 0.6210 | 0.6107 | 0.6734 |
| 4 | 0.6070 | 0.5988 | 0.6147 |
| 5 | 0.5310 | 0.5357 | 0.5624 |
| 6 | 0.5180 | 0.4799 | 0.5120 |
| 7 | 0.4770 | 0.4305 | 0.4727 |
| 8 | 0.4450 | 0.3864 | 0.3644 |
| 9 | 0.3620 | 0.3118 | 0.2926 |
| 10 | 0.3050 | 0.2805 | 0.2676 |
| 11 | 0.2350 | 0.2261 | 0.1558 |
| 12 | 0.2200 | 0.2030 | 0.1444 |
| 13 | 0.1930 | 0.1461 | 0.1335 |
| 14 | 0.1630 | 0.1307 | 0.1230 |
| 15 | 0.1560 | 0.1041 | 0.1033 |
| 16 | 0.1400 | 0.0730 | 0.0940 |
| 17 | 0.1310 | 0.0570 | 0.0767 |
| 18 | 0.1050 | 0.0328 | 0.0527 |
| 19 | 0.0720 | 0.0183 | 0.0455 |
| 20 | 0.0590 | 0.0109 | 0.0320 |
| 21 | 0.0330 | 0.0090 | 0.0257 |
| 22 | 0.0240 | 0.0073 | 0.0197 |
| 23 | 0.0120 | 0.0047 | 0.0140 |
| 24 | 0.0090 | 0.0009 | 0.0085 |
| 25 | 0.0050 | 0.0003 | 0.0048 |
| 26 | 0.0030 | 0.0000 | 0.0015 |

# 6. CONCLUSION

In the presented work, an effort has been made to introduce a new technique that helps to increase the probability of detection for negative values of SNR, and also decreasing the probability of false alarm.

It has been observed from the above stated simulation that when the squaring operation is replaced with the cubing operation in the proposed detector, the probability of missed detection lowers down; which is indirectly proportional to the probability of detection, hence showing that probability of detection is improving with the proposed method.

The proposed method demonstrates evidently better results showing that the probability of missed detection and false alarm decrease as the improved method is used. It shows a noticeable decrease in probability of missed-detection especially for the negative SNR values which was the main motive of the approached work. Hence the results show that the proposed technique is a colossal improvement over the conventional method.

**Authors**

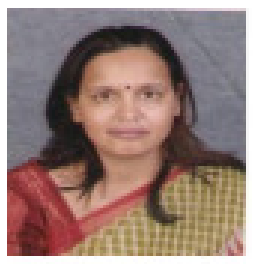

S.Taruna is an active researcher in the field of communication and mob ile network. She is currently working as assistant professor in department of computer science at Banasthali University, Jaipur, India. She has done M.Sc from Rajasthan University and her PhD from Banasthali University, Jaipur, India. She has presented many papers in National and International conferences.

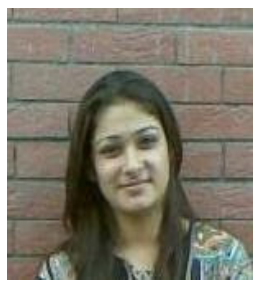

Bhumika Pahwa received her B.Tech degree in Information Technology from Manav Rachna College of Engineering, Faridabad in 2012. She is currently pursuing her M.Tech in IT from Banasthal i Vidyapith. She has published and presented five papers in peer reviewed journals and National conferences. Her research interest includes Cognitive Radio, wireless communication system and other wireless technologies.